\documentstyle{article}\textheight53pc\textwidth6.5in
\begin{document}
 \newcommand{\w}{\omega} \newcommand{\E}{\mbox{E}} \newcommand{\ov}{\overline}
 \newcommand{\qed}{{\rule{1em}{1.5ex}}}\newcommand{\edf}{\stackrel{\rm def}{=}}
 \def\a{\alpha} \newcommand{\N}{{\mbox{\bf\makebox[0pt][l]{\hspace*{1pt}I}N}}}

\title {Randomness and Non-determinism\thanks {in ``European Summer Meeting
of the Association for Symbolic Logic, 1992''. J. Symb.~Logic,
58(3):1102-1103, 1993.}} \author {Leonid A. Levin\thanks {Supported by NSF
grant CCR-9015276.}
 \\Boston University\thanks {Computer Science department, Boston, MA 02215;
 (e-mail: Lnd bu.edu)}} \date{}\maketitle

\noindent Exponentiation makes the difference between the bit-size of this line
and the number ($\ll2^{300}$) of particles in the known Universe. The expulsion
of exponential time algorithms from Computer Theory in the 60's broke its
umbilical cord from Mathematical Logic. It created a deep gap between
deterministic computation and -- formerly its unremarkable tools -- randomness
and non-determinism. Little did we learn in the past decades about the power of
either of these two basic ``freedoms'' of computation, but some vague pattern
is emerging in relationships between them.  The pattern of similar techniques
instrumental for quite different results in this area seems even more
interesting.  Ideas like multilinear and low-degree multivariate polynomials,
Fourier transformation over low-periodic groups seem very illuminating.  The
talk surveyed some recent results. One of them, given in a stronger form than
previously published, is described below.

$|x|$ will denote the length of string $x$. Let P be the set of fast, i.e.\
computable in time $T_{f(x)}=|x|^{O(1)}$, algorithms $f(x)$ on binary strings.
[Blum Micali 82, Yao 82] proposed a fast deterministic way to generate ``nearly
perfect'' randomness, using the idea of a {\em hard core} or {\em hidden} bit.
They assume certain length preserving functions $f\!\in$P to be one-way (OWF),
i.e.\ infeasible to invert (a non-deterministically easy task). Suppose it is
hard to compute from $f(x)$ not only $x$ but even its one bit $b(x)\in\{\pm1\},
\ b\!\in$P. Moreover, assume that even guessing $b(x)$ with any noticeable
correlation is infeasible.  If $f$ is bijective, $f(x)$ and $b(x)$ are both
random and {\em appear to be} independent to any feasible test, thus increasing
the initial amount $|x|$ of randomness by one bit.  Then, a short random seed
$x$ can be transformed into an arbitrary long string $\a(1),\a(2),\ldots$: \
$\a(i)= b(f^{(i)}(x))$. Such $\a$ passes any feasible randomness test.
[Goldreich Levin 89] showed that every OWF $f$ has such a hidden bit with
security of $f$ and $b$ polynomially related. It also gives more details on the
definitions below.  Here this result is strengthened to yield the same security
for $f$ and $b$.

Let $\ov P$ be the set of probabilistic algorithms $A(x,\w)$ using coin-flips
$\w\in\{0,1\}^\N$ and running in average over $\w$ time $\E_\w T_{A(x,\w)}=
|x|^{O(1)}$. An {\em inverter} $I\in\ov P$ for $f$ attempts to compute from
$f(x)$ a list of strings containing $x$. Its success rate $s_{I,f}(n)$ is the
probability of $\{x\in \{0,1\}^n,\w\!: x\in I(f(x),\w)\}$. A {\em guesser} for
$b:S\to\{\pm1\}$ on $f\!\in$P is a $\ov P$-algorithm $G(y,\w)\in\{0,\pm1\}$.
Its success rate is $s_{G,b,f}(n)=(\E_{x,\w}G(f(x),\w)b(x))^2/\E_{x,\w}
G(x,\w)^2$, i.e.\ the inverse sample size needed to notice the correlation with
$b$.  The {\em security} of OWF $f$ or of its hidden bit $b$ is a lower bound
of $1/s(n)$ for all $I$ (or $G$) and big enough $n$.

Let us pad a OWF $f$ to $f'(x,r)=(y,r)$, $y=f(x)$; $x,y,r\in Z^n_2$.
 Let $b(x,r)=(-1)^{(x\cdot r)}$; $v_i=0^{i\!-\!1}10^{n-i}$. We fix $y,\w$,
denote $G_r=G(y,r,\w)$ and $c(x)=\E_r b(x,r)G_r/\sqrt{\E_rG^2_r}$.
We will build an inverter for $f(x)$ with the success probability $\ge c(x)^2$.
 Due to Cauchy-Schwarz inequality, its overall success rate $\ge s_{G,b,f}$.

Note that $c(x)$ (if extended to real vectors) is a generic, up to a constant
factor, multilinear function with coefficients given by $G_r$.
 It is the Walsh (Fourier over group $Z^n_2$) transformation of $G_r$.

Say $c(x)>0$. Then averaging $(-1)^{(x\cdot r)} G_r$ over $>2n/c(x)^2$ random
pairwise independent $r$ yields $>0$ with probability $>1-1/2n$, and the same
for $(-1)^{(x\cdot[r+v_i])} G_{r+v_i}= (-1)^{(x\cdot r)} G_{r+v_i} (-1)^{x_i}$.
Let $k> \log(2n/c(x)^2)$. Take a random matrix $R\in \{0,1\}^{n\times k}$. Then
the vectors $Rp$, $p\in\{0,1\}^k\setminus\{0^k\}$ are pairwise independent.
So, for a fraction $\ge1-1/2n$ of $R$, sign$\sum_p (-1)^{xRp} G_{Rp+v_i}=
(-1)^{x_i}$. We could thus find $x_i$ for all $i$ with probability $1/2$ if we
knew $z=xR$. But $z$ is short: we can try all $2^k$ possible values!

So, the inverter flips $l(\w)\le2n$ coins until the first $0$ and sets $k=l+
\lceil\log5n\rceil$. With $2c(x)^2$ chance $k$ is large enough. Then for a
random $R$ and all $i,p$ it computes $g_i(p)=G_{Rp+v_i}$. It uses Fast Fourier
on $g_i$ to compute $h_i(z)=\sum_p(-1)^{(z\cdot p)}g_i(p)$. The sign of
$h_i(z)$ is the $i$-th bit for the $z$-th member of output list. \qed

Using an $n\times i$ Toeplitz matrix in place of the vector $r$ one can extract
$i$ bits from $x$ rather than one. According to [GL], this will decrease the
security of the bits by a factor of $2^i$.

The power of the above theorem (and a weaker one in [GL]) can be seen even in
the trivial case $f(x)=0$. It is an OWF if $x$ has any distribution such that
the probability of $x$ is always, say, $<4^{-i}$. No relation between $|x|$ and
$i$ or other condition is needed. Such ``junk'' $x$ are much more available
than random uniformly distributed strings.  Having a fixed random $r$ and an
unlimited supply of such $x$, one can keep extracting $i$ ``nearly perfect,''
with security $2^i$, random bits from each $x$. In this case the security (of
$f$ and thus of $b$) is probabilistic: it holds for functions i.e.\ algorithms
with any oracle.  This method requires no additional proof and puts much weaker
assumptions on the distribution than the original Vazirani result.

The hidden bit works for any OWF. But only ``almost bijections'' are known to
yield pseudorandom generators without crucial security loss. Suppose, however,
we have a length preserving $f\in$P with a polynomial fraction of $y$ for which
$x\in f^{-1}(y)$ is hard to find. We may try to convert it into an ``almost
bijection'' with the same property. It may be that $f'(a,x)= (a, f(x)+ax)$
(where $a$ is in a finite field and slightly longer than $x$) will always do.

\section*{Acknowledgements} The idea of using pairwise independent strings and
 vectors $v_i$ is due to [Alexi Chor Goldreich Schnorr 84]. Its use to modify
[GL] proof was suggested to me by Charles Rackoff and R. Venkatesan. Avi
Wigderson agreed to explain to me the Fast Fourier Transformation over
non-cyclic groups, like the Walsh transformation over $Z^n_2$.  I am indebted
to them for this crucial information.


\begin{thebibliography}{9}
	 \bibitem[ACGS]{ACGS} W.Alexi, B.Chor, O.Goldreich and C.P.Schnorr.
 RSA and Rabin Functions: Certain Parts Are As Hard As the Whole.
 {\em SIAM J. Comput.}, 17:194-209, 1988; also {\em FOCS,} 1984.
	\bibitem[BFLS]{BFLS} L.Babai, L.Fortnow, L.Levin, M.Szegedy.
 Checking computations in polylogarithmic time.\\
 {\em ACM Symp.\ on Theory of Computing}, pp.~21-31, 1991.
	 \bibitem[BM]{BM} M. Blum, S. Micali.
 How to Gen\-er\-ate Cryptographically Strong Sequences of Pseudo-Random Bits.
{\em SIAM J. Comput.}, 13:850-864, 1984; also {\em FOCS}, 1982.
	\bibitem[GL]{GL} O.Goldreich, L.Levin. A Hard-Core Predicate for all
One-Way Functions.\\ {\em ACM Symp.\ on Theory of Computing}, pp.~25-32, 1989.
	\bibitem[V]{V}U.Vazirani. Efficiency Considerations in Using
Semi-random Sources.\\  {ACM Symp.\ on Theory of Computing}, pp.~160-168, 1987.
       \bibitem[Y]{Y}A.C. Yao. Theory and Applications of Trapdoor Functions.\\
{\em Proc.~of IEEE Symp.~on Foundations of Computer Sci.}, pp.~80-91, 1982.

\bibitem[add]{add} additional references (in ICM-94, not here):\\
O. Goldreich, S. Goldwasser, S. Micali.  How to Construct Random
Functions.\\ FOCS-84.  {\em J. ACM}, 33/4:792-807, 1986.\\
 L. Levin.  One-Way Functions and Pseudorandom Generators.\\
 {\em Combinatorica}, 7(4):357-363, 1987. (earlier in STOC-1985).\\
 Johan Hastad, Russell Impagliazzo, Leonid A. Levin, Michael Luby.\\
 {\em Construction of a Pseudo-Random Generator from any One-Way Function.}\\
 Internat.Comp.Sci.Inst. (Berkeley).  Tech.Rep. 91-068, pp.1-36. 12/1991.
 To appear in SICOMP.\\ Earlier versions in STOC-1989 [ILL] and STOC-1990 [H].
	\end{thebibliography}
  \end{document}